\documentclass[aip,jap,reprint,amsmath,amssymb]{revtex4-1}
\usepackage{amsmath}
\usepackage{graphicx}
\usepackage{dcolumn}

\begin{document}
\title{Magnetic field induced shell-to-core confinement transition in type-II semiconductor quantum wires}
\author{R. Mac\^edo}
\author{J. Costa e Silva}
\affiliation{Departamento de Ci\^{e}ncias Exatas e Naturais,
Universidade Federal Rural do Semi-\'Arido, 59600-900 Mossor\'o, RN, Brazil}
\author{A. Chaves}
\author{G. A. Farias}
\affiliation{Departamento de F\'isica, Universidade Federal do
Cear\'a, CP 6030, Campus do Pici, 60455-900 Fortaleza,
Cear\'a, Brazil}
\date{ \today }
\author{R. Ferreira}
\address{Laboratoire Pierre Aigrain, Ecole Normale Superieure,
CNRS UMR 8551, Universit\'e P. et M. Curie, Universit\'e Paris
Diderot, 24 rue Lhomond, F-75005 Paris, France}

\begin{abstract}
We investigate the excitonic properties of a core-multishell semiconductor nanowire with type-II band mismatch, i.e. with spatially separated electrons and holes, under an external magnetic field. Our results demonstrate that, depending on the core wire radius, the carrier in the type-II band exhibits either a quantum dot-like or a quantum ring-like energy spectrum, corresponding to a carrier confinement in the core wire or in the outer shell, respectively. In the latter, a shell-to-core confinement transition can be induced by increasing the magnetic field intensity, which may lead to interesting photocurrent properties of these confining structures, tunable by the external field.
\end{abstract}

\maketitle

\section{Introduction}

The current need for smaller and more efficient electronic devices has stimulated the development, fabrication, and investigation of novel semiconductor-based low-dimensional confining structures, such as quantum wells, wires and dots. Over the past few years, the growth of core-shell \cite{Coreshell} and longitudinally heterostructured \cite{SLNW} quantum wires composed by a variety of different semiconductor materials have been reported in the literature. \cite{D. Appell, M. Law} In particular, recent articles \cite{Lauhon, Ishai} have shown the growth of Si/Si$_{1-x}$Ge$_x$
core-multishell wires, using the CVD (chemical vapor deposition)
and VLSD (vapor-liquid-solid deposition) growth methods. These structures show excellent transport properties, making possible their application in high-mobility devices. \cite{H.-J. Choi} Another interesting feature of the Si/Si$_{1-x}$Ge$_x$
heterostructure is the fact that, as inferred from experimental results, \cite{1} and according to
electronic affinity calculations, \cite{2} this heterostructure exhibits a type-II band alignment, where the conduction
(valence) bands mismatch between materials is such that the
Si$_{1-x}$Ge$_x$ layer acts as a barrier (well) region for electrons (holes), specially for Ge concentrations $x > 0.3$. \cite{3, Yang} Nevertheless, it has been demonstrated that
type-I alignment in these structures can be achieved as well through
strain engineering, \cite{Wang, AndreyWell} which is important for
the application of such heterostructure materials in
photoluminescence devices. 

In this paper, we present a theoretical study of the excitonic
properties of a core-multishell (CMS) type-II quantum wire,
consisting of a Si core covered by a Si$_{0.7}$Ge$_{0.3}$ internal
shell and a Si external shell, \cite{Lauhon} where the hole is confined at the
Si$_{0.7}$Ge$_{0.3}$ shell, while the electron must be localized
in one of the Si regions of the system. \cite{3} In this case, it is
unclear whether the electron wavefunction would be mostly
distributed in the external Si shell, in the Si core, or in both. As we will demonstrate further,
our results show that the electron is mostly localized in the
external Si shell for small values of the core radius, while for
larger core radii, it localizes in the Si core. In the former
case, Aharonov-Bohm (AB) oscillations are observed in the exciton energy for lower
magnetic field intensity, whereas higher values of the field
induce a shell-to-core transition of the electron confinement, so
that the AB oscillations cease to occur. Such a transition may
have an important effect, for example, on the electrons mobility along the core
wire, since this quantity depends on the electron probability
density.

\section{Theoretical Model}

In order to simplify the calculations, we have considered the following reasonable approximations: (i) we will focus on the case of a circularly symmetric cylindrical core wire, surrounded by multiple shells. \cite{Lauhon} In fact, core-shell quantum wires with hexagonal cross-section have also been experimentally observed. \cite{Guo, Mohan} Previous papers in the literature have theoretically demonstrated that the main effect of such hexagonal geometry was an increase of the carriers concentration around the corners of the hexagonal confining shell. \cite{Ferrari} Similar effect is also observed for an elliptic shell, where carriers concentrate in the regions of higher curvature; \cite{CostaeSilvaSSC, Submitted} (ii) Heavy hole-Light hole coupling and band mixing effects are neglected - this is an usual and fair approximation \cite{Barbagiovanni, Slachmuylders} that simplifies a lot the calculations, by avoiding the need to diagonalize a valence band matrix, which is a very time consuming procedure; (iii) The dielectric mismatch effect between the different layers and between the outmost shell and the vacuum are neglected. In fact, the former is not expected to have a significant contribution, since this effect is proportional to $1-\epsilon_1/\epsilon_2$, \cite{Sousa} where $\epsilon_{1}$ and $\epsilon_{2}$ are the dielectric constants of the Si core and Si$_{0.7}$Ge$_{0.3}$ shell, respectively, which are very similar. The latter could lead to significant corrections, but it has been demonstrated that dielectric mismatch effects become less important for large radii, \cite{Slachmuylders} and all the results investigated here are obtained for a very large width of the outmost shell, so that this dielectric mismatch effect is also negligible; (iv) Different papers in the literature have reported that a combination between strain and quantum confinement effects in Si/Si$_{1-x}$Ge$_x$ heterostructures leads to a type-II band alignment formed by the $\Delta_2$ electron and the heavy hole band edges. \cite{3, Wang} We assume that our radial Si/Si$_{0.7}$Ge$_{0.3}$ heterostructure is in this type-II regime, considering the same carriers effective masses and band offsets as found in Refs. \onlinecite{3, Yang, Wang, CostaeSilvaQWR}.

Using the cylindrical coordinates system, we take the $(\rho,\theta)$-plane as the confinement plane and $z$ as the free direction. Assuming the symmetric gauge for the vector potential, the exciton Hamiltonian, within the effective mass approximation,
reads \cite{F. M. Peeters, CostaeSilvaQWR}
\begin{eqnarray}
H_{exc} = H_e + H_h - \frac{1}{\mu_{\perp}} \frac{\partial^2}{\partial z^2}  - \frac{2}{\epsilon\vline \vec{r}_e - \vec{r}_h \vline}
 \label{eq2.1},
\end{eqnarray}
where the indexes $e$ and $h$ stand for electron and heavy hole,
respectively, and $H_e$ ($H_h$) is the electron (hole) single particle Hamiltonian
\begin{equation}
H_i = \left\{- \frac{1}{m_i^{\parallel}}\left(\vec{\nabla}_{2D,i} - \frac{2\pi}{\Phi_0}\vec A\right)^2
+ V_i(\rho_i)\right\} 
\end{equation}
where $m_i^{\parallel}$ is the in plane effective mass of each charge
carrier, $\Phi_0 = h/e$ is the magnetic quantum flux, $\mu_{\perp}$ is the electron-hole reduced mass in the $z$-direction, and $z = \mid z_e - z_h \mid $ is the
electron-hole relative coordinate. The $V_{i}(\rho_i)$ term is the circularly symmetric hetero-structure potential, due to the bands
mismatches, whereas the last term of Eq. (\ref{eq2.1}) accounts for the electron-hole Coulomb interaction. Energy
variables are divided by the Rydberg energy $Ry$, spatial
variables by the Bohr radius $a_0$, and effective masses by the
free electron mass $m_0$, in order to make the equations
dimensionless. Assuming a Si core wire of radius $\rho_1$, surrounded by a Si$_{0.7}$Ge$_{0.3}$ shell of width $W = \rho_2 - \rho_1$, whose interface with the external Si shell is placed at $\rho_i = \rho_2$, the heterostructure potential in the conduction (valence) is given by $V_e (\rho_e) = 0$ $[V_h (\rho_h) = V_{0h}]$ if $0 < \rho_i  < \rho_1$ or $\rho_i > \rho_2$, and $V_e (\rho_e) = V_{0e}$ $[V_h (\rho_h) = 0]$ otherwise.

In order to solve the Schr\"odinger equation for the exciton $H_{exc}\Psi = E_{exc}\Psi$, we neglect initially the coulombic coupling, so that the solutions can be put in the form $\Psi = \psi_e(\vec \rho_e)\psi_h(\vec \rho_h)\psi_z(z)$ to obtain $H_i\psi_i = E_i\psi_i$
for each carrier $i$, whose in-plane position vector is given by
$\vec{\rho_i}$. Due to the type-II alignment in such a structure, the hole is localized in the Si$_{1-x}$Ge$_x$ shell, whereas the electron alone is no longer confined by the band mismatch of the wire materials, but can become laterally bounded by the
Coulomb interaction, as we discuss in the following. Indeed, the single
electron solutions of $H_e\psi_e = E_e\psi_e$ form a continuum. Therefore, in order to solve the type-II problem, we first numerically \cite{F. M. Peeters} solve $H_h\psi_h = E_h\psi_h$ to obtain the ground hole state wavefunction $\psi(\rho_h)$. The electron state is then calculated by considering the hole motion "frozen" in its ground state. This leads to the following equation for the electron eigenproblem:
\begin{equation} \label{eq.tosolve}
\left[H_e + E_h + V_{eff}(\rho_e)\right]\psi_e (\rho_e)= E_{x} \psi_e (\rho_e),
\end{equation}
where $V_{eff}(\rho_e) = \langle\psi_h\psi_z|H_{exc}|\psi_h\psi_z\rangle$ is the effective Coulomb potential. By assuming a gaussian form for the relative motion along the wire axis \cite{Frank L.
Masdarasz} 
\begin{equation}\label{eq.Gaussian}
\psi_z(z) = \frac{1}{{\sqrt \eta  }}\left( {\frac{2}{\pi }}
\right)^{1/4} \exp \left( { - \frac{{z^2 }}{{\eta ^2 }}} \right),
\end{equation}
one obtains
\begin{equation} \label{eq.effectivepotelectron}
V_{eff}(\rho_e) = \frac{1}{{\mu_\bot \eta ^2 }} - \frac{2}{\varepsilon \eta }\sqrt {\frac{2}{\pi }} \int\limits_{V_h} {\left|{\psi_h}\right|^2}\exp(a)K_0(a/2) d\vec \rho_h,
\end{equation}
where $a = \left| \overrightarrow {\rho_e} - \overrightarrow{\rho_h} \right|^2/\eta^2$, $K_0(x)$ is the modified zero-order
Bessel function of second kind, $\eta$ is a variational parameter that minimizes the exciton energy, and the second term is integrated only in the hole space. Notice that the influence of the electron density on the distribution of the hole is neglected, but this is not supposed to have a significant influence on the results, since the hole is strongly bound in the shell by the bands mismatch potential. It is also important to stress out that Eq. (\ref{eq.tosolve}) was solved by a variational procedure using the gaussian function Eq. (\ref{eq.Gaussian}) for the relative motion in the axial direction. This simplification is often used, \cite{CostaeSilvaQWR} but is known that it provides slightly overestimated values for the ground state energy, which can be further quantitatively improved by using more sofisticated variational functions. \cite{Harrison, Caetano} Nevertheless, the qualitative features of the spectra, such as the shell-to-core transitions discussed in the following sections, which are the main focus of this paper, are not affected by choosing different variational functions for the axial direction.

\section{Results and Discussion}

\begin{figure}[!h]
\centerline{\includegraphics[width=0.8\linewidth]{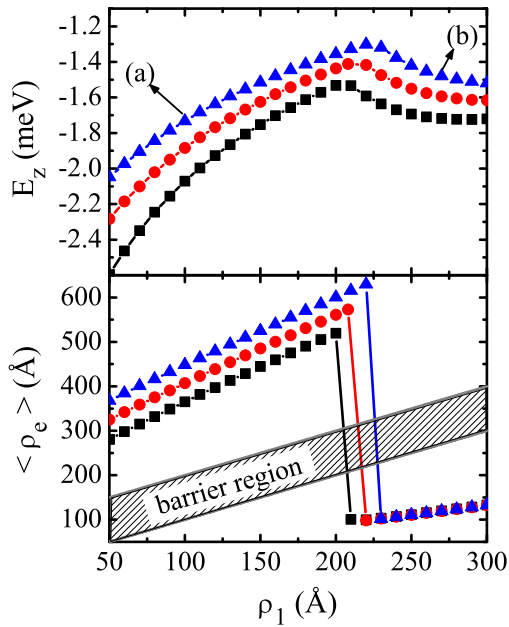}}
\caption{(color online) Binding energy (top) and electron average radius (bottom)
as a function of the core radius, considering
the confining shell width as $W$ = 50 \AA\, (black squares), 75
\AA\, (red circles) and 100 \AA\, (blue triangles). The arrows indicate the points $\rho_1$ = 100 \AA\, and 270
\AA\,, whose wave functions are illustrated in Fig. \ref{fig:2}
(a) and (b), respectively.} \label{fig:BindingAverageTypeII}
\end{figure}

\begin{figure}[!h]
\centerline{\includegraphics[width=0.8\linewidth]{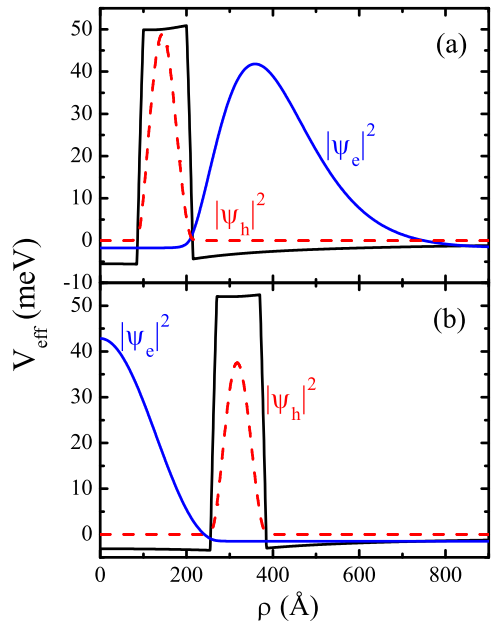}}
\caption{(color online) Electron effective potential (black solid), considering $\rho_1$ = 100 \AA\, (a) and 270 \AA\, (b), along with
their respective electron (blue solid) and hole (red dashed) wave functions.}
\label{fig:2}
\end{figure}

As we mentioned previously, the type-II Si/Si$_{0.7}$Ge$_{0.3}$/Si CMS quantum wire exhibits
an interesting feature: the electron can be
confined either in the external Si shell, or in the internal Si
core. These two situations are expected to lead to completely
different excitonic behaviors. Indeed, the dependence of the
electron-hole binding energy on the inner core radius $\rho_1$, for three
values of the Si$_{0.7}$Ge$_{0.3}$ shell width $W$, is
illustrated in Fig. \ref{fig:BindingAverageTypeII} (top), where one
easily observes that as the core radius increases from 50 \AA\,,
the binding energy increases only until a critical value of this
parameter is reached. After this value of $\rho_1$, the energy starts
to decrease. Concomitantly, as shown in the lower
panel of Fig. \ref{fig:BindingAverageTypeII}, the electron average radius $\langle \rho_e \rangle$ exhibits an abrupt transition exactly at the critical
core radius, where the electron changes its radial average
position from the outer shell towards the central core. Finally, this
intriguing shell-to-core transition with increasing inner radius is
confirmed by Figs. \ref{fig:2} (a) and (b), where the electron (blue solid)
and hole (red dashed) lateral probability amplitudes are presented, along with the effective electron
confinement potential, for core radii $\rho_1$ = 100 \AA\, (a) and 270 \AA\, (b), considering a shell width $W$ = 100 \AA\, (arrows in the upper panel of Fig. \ref{fig:BindingAverageTypeII}). We clearly see that the electron wave function jumps from the external Si shell towards the Si core as the core radius increases.

Such a shell-to-core transition simply reflects the competition
between shell and core states for electrons. In order to clarify
this trend, it is worth noticing two points:  (i) First, that the
independent electron posses a quasi-bound state, which is
strongly localized in the core region and whose energy varies as
1/$\rho_1^2$ (other states have only a small presence in the core).
(ii) Second, the hole is tightly bound to the SiGe
layer, forming a thin ring of positive charge enveloping the
core, and generating thereby an attractive effective potential
for the electron (Eq. (\ref{eq.effectivepotelectron})) that is sensibly higher in the Si core than in the external Si layer. The coulombic binding of the resonant
electron state with the positive cloud decreases much slowly with
$\rho_1$ (roughly as 1/$\rho_1$). Thus, for small inner radius $\rho_1$ the
electron resonance is too high in energy. However, as the thickness
of the core radius increases, the resonant state energy decreases
very fast and, account to the coulombic coupling, the core
confinement of the electron becomes more energetically favorable
as compared to the much weaker binding of outside electron
states.
 
\begin{figure}[!h]
\centerline{\includegraphics[width=0.9\linewidth]{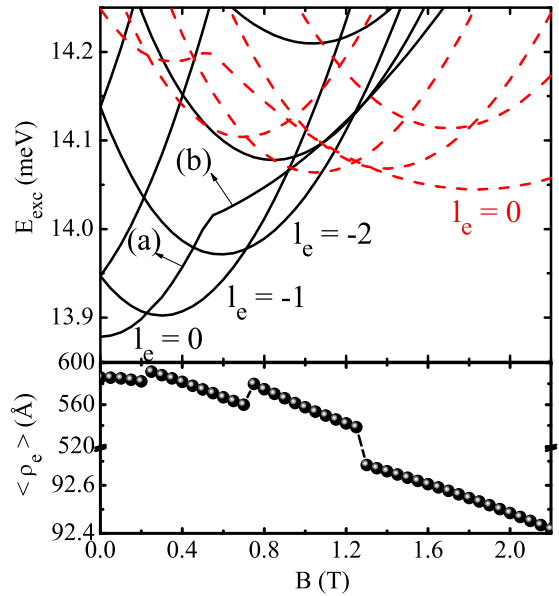}}
\caption{(color online) Exciton transition energy (top) and ground state electron average radius (bottom) as a function of the
magnetic field intensity, for $W = 100$ \AA\, and $\rho_1$ = 200 \AA\,. We
consider two values for the hole angular momentum index, $l_h = 0$
(black solid) and $l_h = -1$ (red dashed), and several values for
the electron angular momentum index $l_e$. The wave functions for the $l_h = 0$ and $l_e = 0$ states
at the two values of magnetic fields pointed by the arrows are
illustrated in Fig. \ref{fig:TypeIImagWaves}. Notice the vertical axis of the bottom panel has a break between 92.8 \AA, and 520 \AA\,, in order to improve the visualization of the curve.}
\label{fig:TypeIImagEn}
\end{figure}

The electron confinement in the external shell can lead to rather
interesting features in the presence of an applied magnetic field
parallel to the wire axis. As the magnetic field intensity
increases, the hole confinement energy is expected to present AB
oscillations, since this carrier is strongly confined within the internal
shell, and such a shell confinement shares the same properties as the in-plane motion of a particle in a quantum ring. In fact, the AB effect for a carrier confined in the shell region of a core-shell wire has been experimentally confirmed by the
observation of quantum interference effects on the
magnetoresistance of a In$_2$O$_{3}$/InO$_x$ nanowire. \cite{Jung} As the electron is in the external shell for small core
radii, one can expect that its energy will also exhibit AB
oscillations, although with a different period, since its
average radius differs from that of the hole, for they are
confined in different shells. However, the magnetic field pushes
the electron towards the center, which is also a Si layer, thus,
for high magnetic field intensities, a transition from the
electron confinement at the external layer to the Si core is also
expected to occur and, once the electron is in the core for higher
magnetic fields, its energy cannot exhibit AB oscillations
anymore. This is illustrated by Fig. \ref{fig:TypeIImagEn}, which
shows the exciton energy $E_{exc}$ (top) and the ground state electron average radius (bottom) as a function of the magnetic
field intensity $B$. The results are obtained for $W = 100$ \AA\, and $\rho_1 = 200$ \AA\,, so that the system is close to the configuration where the core confinement is more energetically favorable (see Fig. \ref{fig:BindingAverageTypeII}), which helps one to obtain the magnetic field induced shell-to-core confinement transition with lower magnetic fields. Notice that such a confinement transition for the electron does not have an analog in the context of semiconductor quantum rings, since these rings are normaly obtained by a volcano-like structure surrounded by a layer of a different material, \cite{Lorke} so that a type-II band alignment would lead to a carrier wave function surrounding the whole ring surface. \cite{Timm} Nevertheless, the tunability of AB oscillations in type-I quantum rings have been also recently reported, but due to a different confinement effect. \cite{BinLi}

\begin{figure}[!h]
\centerline{\includegraphics[width=0.9\linewidth]{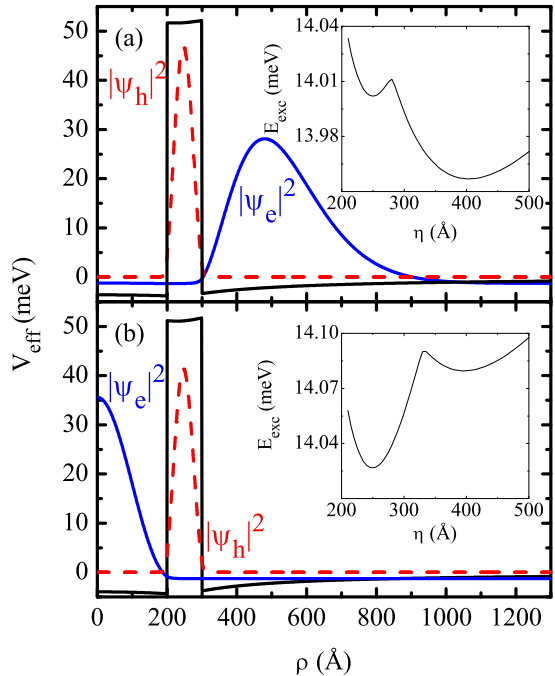}}
\caption{(color online) Electron effective potentials (black), along with their
respective electron (blue) and hole (red) wave functions, for the
two different values of magnetic field pointed in Fig.
\ref{fig:TypeIImagEn}: (a) $B$ = 0.4 T and (b) $B$ = 0.65 T. The
insets show the exciton transition energy as a function of the
variational parameter $\eta$ for each value of magnetic field.}
\label{fig:TypeIImagWaves}
\end{figure}

The kinks in the curves corresponding to the $(l_e, l_h) =
(0, 0)$ and $(0, -1)$ energy states at $B \approx$ 0.55 T are direct consequence
of such a magnetic field induced shell-to-core transition. Indeed,
by analyzing the squared modulus of the electron wave functions
for this system in Fig. \ref{fig:TypeIImagWaves}, one observes
that for $B = 0.4$ T (a), the $l_e = 0$ electron state,
considering $l_h = 0$, is confined at the Si external shell,
whereas a core confinement is observed for this state at $B =
0.65$ T (b). Similar results are observed for $l_h = -1$ and $l_e
= 0$. More details on this transition are given in the insets of
Fig. \ref{fig:TypeIImagWaves}, which show the exciton energy as a
function of the variational parameter $\eta$ in Eq.
(\ref{eq.effectivepotelectron}). Two local minima are observed, where the one for
lower (higher) $\eta$ represents an electron state confined in the
core (shell). The global minimum switches from one to the other as
the magnetic field increases, leading to the core-to-shell
transition illustrated in Figs. \ref{fig:TypeIImagEn} and
\ref{fig:TypeIImagWaves}. 

Notice, however, that the shell-to-core transition represented by
the kinks in Fig. \ref{fig:TypeIImagEn} (top) does not occur for the
exciton ground state, which for $B = 0.55$ T is the $(l_e, l_h) =
(-1, 0)$ state, that is still confined at the external shell. This would make such a transition harder to be experimentally observed. On the other hand, a shell-to-core transition occurs for the exciton ground state at a higher magnetic field intensity, $B \approx 1.2$ T. This is clearly evidenced in Fig. \ref{fig:TypeIImagEn} (bottom), where the ground state electron average radius oscillates around $\approx$ 570 \AA\, (outer shell region) for lower magnetic fields, but ceases to oscillate for $B$ higher than $\approx 1.2$ T, when its value decreases to $\approx$ 92.6 \AA\, (core region).
Indeed, due to the shell confinement of the electron for smaller magnetic
fields, angular momentum transitions and AB oscillations occur as
the magnetic field increases. The period of these oscillations
($\approx$ 0.37 T) is in good agreement with the value obtained by
a simple model of the angular energy states of the system
\cite{CostaeSilvaSSC}, which predicts a period of $\Phi_0 = h/e$ for
the magnetic flux, if one considers a magnetic flux through a
shell with radius given by the electron average radius $\langle
\rho_e \rangle$ $\approx$ 600 \AA\, shown in Fig.
\ref{fig:BindingAverageTypeII} for this system. However,
for $B \gtrsim 1.2$ T, the $(l_e, l_h) = (0, -1)$
becomes the ground state and these electron angular momentum transitions
cease to occur, as a consequence of the fact that the electron in this exciton state for such a field is confined in the Si core, with $\langle \rho_e \rangle$ $\approx$ 92.6 \AA\,. Once this core
confined state becomes the ground state, AB oscillations start to occur with a much higher period, being present only due to the contribution of the hole states: since the hole is always confined in the Si$_{0.7}$Ge$_{0.3}$ shell, with average radius 250 \AA\,, it is expected to exhibit AB oscillations with a period of $\approx$ 2.11 T. In fact, the hole angular momentum transition from $l_h =$ 0 (black solid) to $l_h$ -1 (red dashed) in the ground state exciton for $B \approx 1.05$ T observed in Fig. 3 (top) is just a consequence of the AB effect, which predicts angular momentum transitions at half integers of the AB period. Other transitions would occur at higher magnetic fields, out of the range shown in Fig. 3. Therefore, the transition from lower to higher periods of the AB oscillations of the exciton ground state energy at a certain value of magnetic field in a CMS type-II quantum wire provides a clear evidence of the shell-to-core confinement transition, which can be verified by future experiments.

This shell-to-core transition may lead to interesting photocurrent properties. Under illumination, photogenerated holes end up in the Si$_{1-x}$Ge$_{x}$ shell and either become trapped or conduct with low mobility. In both cases they form a positive radial shell that retains electrons nearby. For small core radius, electrons flow in the outside region, and this quasi-onedimensional system can be visualized as a ``wrapped'' analogous of standard quasi-twodimensional heterojunctions. For bigger core radius, on the contrary, electron transport takes place in the inner core region with an eventually rather different mobility along the core wire axis. Our results (see e.g. Fig. \ref{fig:TypeIImagEn}) suggest that an axial magnetic field can be used to tune from one regime to the other. The study of such transport properties is however beyond the scope of this work.

\section{Conclusions}

In summary, we demonstrate that in a type-II structure consisting of a Si
core wire surrounded by a single shell Si$_{0.7}$Ge$_{0.3}$ and
covered by an external Si shell, the electrons may be confined
either in the core or in the outmost shell, depending on the
radius of the core region. In the latter case, a shell-to-core
transition for the electrons confinement can be induced by an
external magnetic field, even for the exciton ground state, which
is suggested as a way to tune the mobility along the core wire
axis.

\acknowledgements
This work has received financial support from
the Brazilian National Research Council (PRONEX-CNPq), and Funda\c{c}\~ao
Cearense de Apoio ao Desenvolvimento Cient\'ifico e Tecnol\'ogico (Funcap).

\end{document}